\documentclass[12pt]{article}
\usepackage{graphicx}

\newsavebox{\sboxpubnumber}
\newsavebox{\sboxpubdate}
\newcommand{\pubdate}[1]{\begin{lrbox}{\sboxpubdate}{#1}\end{lrbox}}
\newcommand{\pubnumber}[1]{\begin{lrbox}{\sboxpubnumber}{\begin{tabular}{l} #1 \\
				 \usebox{\sboxpubdate}
				 \end{tabular}}
                           \end{lrbox}
                           \pubblock}
\newcommand{\Title}[1]{\begin{center} {\Large #1 } \end{center}}
\newcommand{\Author}[1]{\begin{center}{ \sc #1} \end{center}}
\newcommand{\Address}[1]{\begin{center}{ \it #1} \end{center}}
\newcommand{\andauth}{\begin{center}{and} \end{center}}
\newcommand{\pubblock}{\rightline{
			\usebox{\sboxpubnumber}}}
\newenvironment{Abstract}{\begin{quotation}  }{\end{quotation}}
\newenvironment{Presented}{\begin{quotation} \begin{center}
             PRESENTED AT\end{center}\bigskip
      \begin{center}\begin{large}}{\end{large}\end{center}
      \end{quotation}}

\def\beq{\begin{equation}}
\def\eeq#1{\label{#1}\end{equation}}
\def\eeqn{\end{equation}}

\def\beqa{\begin{eqnarray}}
\def\eeqa#1{\label{#1}\end{eqnarray}}
\def\eeqan{\end{eqnarray}}

\begin{document}

\begin{titlepage}
\pubdate{\today}                    
\pubnumber{DF/IST-13.2001} 

\vfill
\Title{Quintessence with Coupled Scalar Fields}
\vfill
\Author{M. C. Bento\footnote{ We
 acknowledge the partial  support of FCT (Portugal)
under  grant POCTI/1999/FIS/36285.}}
\Address{Departamento de F\'{\i}sica/ CFIF, Instituto Superior T\'ecnico\\
         Av. Rovisco Pais, 1049-001 Lisboa, Portugal }
\vfill
\andauth
\vfill
\Author{O. Bertolami$^1$, N. C. Santos}
\Address{Departamento de F\'{\i}sica, Instituto Superior T\'ecnico\\
         Av. Rovisco Pais, 1049-001 Lisboa, Portugal}
\vfill
\begin{Abstract}
We discuss the dynamics of a quintessence model  involving two coupled
scalar  fields. The model presents two types of solutions, namely solutions that correspond to eternal  and   transient  acceleration of the universe.
 In both cases, we obtain values for the cosmological parameters that
 satisfy current observational bounds as well as the nucleosynthesis
 constraint on the quintessence energy density. 

\end{Abstract}
\vfill
\begin{Presented}
    COSMO-01 \\
    Rovaniemi, Finland, \\
    August 29 -- September 4, 2001
\end{Presented}
\vfill
\end{titlepage}
\def\thefootnote{\fnsymbol{footnote}}
\setcounter{footnote}{0}


\section{Introduction}
 
\def\lsim{\mathrel{\rlap{\lower4pt\hbox{\hskip1pt$\sim$}}
    \raise1pt\hbox{$<$}}}
\def\gsim{\mathrel{\rlap{\lower4pt\hbox{\hskip1pt$\sim$}}
    \raise1pt\hbox{$>$}}}

Recent  measurements of the Cosmic Microwave Background (CMB) anisotropies by
 various experiments \cite{cmbexp} suggest that the Universe is flat, with a
 total energy density that is close to the  critical density. However, there
 is strong observational evidence \cite{otherexp} that matter (baryonic plus
 dark) can only account for about one third of the total matter density. On
 the other hand, Type I Supernovae light-curves  indicate that the Universe is
 accelerating at present \cite{supern}. All these observations can be 
 reconciled  if one assumes  that the
the dynamics  of the  universe at present is dominated by  a negative
pressure dark component, the main candidates being a cosmological constant
and  dark energy or  quintessence \cite{Ratra,Caldwell,Zlatev}; the latter is 
characterized by a time-varying  
equation of state parameter,  $w_Q\equiv p/\rho$,  
approaching a present  value $w_Q<-0.6$.

A commom feature of all quintessence models  presented sofar 
is that the asymptotic
accelerating behaviour of the universe  is driven by the dynamics of a
single field.
There are, however,   several motivations for studying the case of
coupled scalar  fields. Firstly, if one  envisages  to extract  the
potential suitable for describing the universe dynamics from
fundamental particle physics theories, it is  most likely that an ensemble
 of coupled scalar
fields (moduli, axions, chiral  superfields, etc) will emerge, for
instance, from the compactification process or  from the localization of
fields in the brane in multibrane models or from  mechanisms
 for the cancellation of the cosmological constant (see
e.g. \cite{Burgess} and  references  therein). Furthermore, coupled
scalar fields are invoked for various desirable features they exhibit, as
 in the so-called hybrid inflationary models
\cite{Linde,Bento2}  and   in reheating  models
\cite{Kofman, Bertolami2} .
Finally, it has been recently pointed out that an  eternally
accelerating universe poses a challenge for string theory, at least in
its present formulation, since asymptotic states are inconsistent with
spacetimes that exhibit event horizons \cite{Hellerman,Fischler}.
Moreover, it is argued that  theories with  a  stable supersymmetric
vacuum cannot relax into a zero-energy ground state if the
accelerating  dynamics is guided by a single scalar field
\cite{Hellerman,Fischler}. The main argument 
 relies   on  the   fact  that,   in  a
supersymmetric theory, one expects  that the asymptotic behaviour of
the superpotential  is given by  $W(\phi) = W_0 e^{-\alpha  \phi /2}$,
which,  in  order to  ensure  the  positivity  of the  $4$-dimensional
potential  $V(\phi) =  8 \vert  \partial_{\phi}  W \vert^2  - 12
  \vert W^2 \vert$
implies that $\vert  \alpha \vert > \sqrt{6}$. However,  this value is
inconsistent  with  the  requirement  of an  accelerated  Universe  at
present  $\vert  \alpha  \vert = \sqrt{3(1+\omega_{Q0})/2} < 1.5$
\cite{Ratra}   as   data   suggest   that  $\omega_{Q0} < - 0.6$
\cite{Perlmutter1}.   The situation  is different  in the  presence of
fields that do not reach  their minima asymptotically.
Indeed, in  this case, the asymptotic behaviour  of the superpotential
would be  better described by  the function $W(\phi) =  W_0 e^{-\alpha
\phi /2} F(\phi, \psi)$, where  $F(\phi, \psi)$ is a polynomial in the
fields  $\phi$  and  $\psi$ and    the  positivity  condition  then becomes:
$\alpha^2 - 6 +  4 [(\partial_{\phi} F)^2 + (\partial_{\psi} F)^2]/F^2
> 0$.  One can  then easily  see  that, by  a suitable  choice of  the
polynomial $F(\phi, \psi)$, the positivity condition can be reconciled
with the requirement of successful quintessence.

Recently, a 
two-field model has been proposed \cite{BentoBS}, exhibiting a class of solutions where
fields  do  not necessarily  settle in  their
minima at present, thus  evading some of the conclusions
of Refs. \cite{Hellerman,Fischler}, concerning the  stability of the
supersymmetric vacua.
The potential is given by

\beq
V(\phi, \psi)=  e^{-\lambda\phi} P(\phi, \psi)~,
\label{pot}
\eeqn
where
\beq  
P(\phi, \psi)= a~+ (\phi-\phi_0)^2+b~(\psi-\psi_0)^2 + c~\phi(\psi-\psi_0)^2 + d~\psi(\phi-\phi_0)^2~,
\label{pot1}
\eeqn
in units where  $M \equiv (8\pi G)^{-1/2}=\hbar=c=1$. 
 Such potentials  arise in the low-energy
limit of fundamental particle physics theories such as string/M-theory
and brane-world  constructions. The overall negative exponential term
in $\phi$  signals that this could be  moduli type field which has
acquired  an interacting  potential with the $\psi$  field.

\section{Cosmological Solutions}

We consider a spatially-flat  Friedmann-Robertson-Walker  (FRW)
universe containing a perfect  fluid with barotropic equation of state
$p_\gamma=(\gamma - 1) \rho_\gamma$,  where $\gamma$ is a constant, $0
\leq \gamma \leq 2$ (for radiation $\gamma=4/3$ and for dust $\gamma =
1$)   and  two  coupled   scalar  fields   with  potential   given  by
Eq.~(\ref{pot}).   The evolution  equations for  a  spatially-flat FRW
model with Hubble parameter $H\equiv \dot a /a$ are 
\beqa
 \dot{H} & = &
- \,   {1\over2}   \left(   \rho_\gamma   +  p_\gamma   +   \dot\phi^2
+\dot\psi^2\right)      \      ,      \\
\dot\rho_\gamma      &=& -3H(\rho_\gamma+p_\gamma)     \    ,     \\  
\ddot\phi    &=& -3H\dot\phi-{\partial_\phi  V}\  ,\\
\ddot\psi  &=&  -  3H\dot\psi  -{\partial_\psi V}\ ,
\label{fried}
\eeqan
where $\partial_{\phi(\psi)} V  \equiv {\partial V \over \partial
\phi(\psi)}$, subject to the Friedmann constraint

\beq  
H^2  =  {1\over3}   \left(  \rho_\gamma   +  {1\over2}\dot\phi^2
+{1\over2} \dot\psi^2 +V \right) \ ,
\label{Friedmann}
\eeqn
The total energy density of the homogeneous scalar fields is given
by $\rho_Q=\dot\phi^2/2+\dot\psi^2/2+ V(\phi, \psi)$.

We define the dimensionless variables

\beq 
x \equiv  {\dot\phi \over  \sqrt{6}\,H}  \quad ;  \quad y  \equiv
{\sqrt{V} \over  \sqrt{3}\,H} \quad ;  \quad z \equiv  {\dot\psi \over
\sqrt{6}\,H} \quad . 
\eeqn
 
The evolution equations can then be written as an autonomous dynamical
system:
 
\beqa
\label{emx}
x' & = & -3x -  \sqrt{{3\over2}} \partial_\phi \ln (V) y^2 + {3\over2}
x f(x,y,z)\ , \\
\label{emy}
y'  &  =  & \sqrt{{3\over2}}  y  \left(x  {\partial_\phi  \ln (V)}  +  z
{\partial_\psi \ln (V)}\right) + {3\over2} y f(x,y,z) \ , \\
\label{emz}
z' & = & -3z -  \sqrt{{3\over2}} \partial_\psi \ln (V) y^2 + {3\over2}
z f(x,y,z)\ , \\
\phi^\prime &=& \sqrt{6}~x~,\\
 \psi^\prime &=&\sqrt{6}~z~ \ ,
\eeqan
where $f(x,y,z)\equiv \gamma (1 - x^2 - y^2 - z^2)
+ 2 (  x^2 + z^2 )$ and  a prime denotes a derivative  with respect to
the logarithm of the scale factor $N\equiv\log(a/a_0)$, where $a_0$ is
the scale factor at  present.
The constraint equation becomes

\beq
{\rho_\gamma \over 3 H^2} + x^2 + y^2 +z^2 = 1 \ .
\label{cons}
\eeqn
and, therefore

\beq 
\Omega_Q\equiv {\rho_Q \over 3H^2} = x^2 + y^2 +z^2\ .
\label{omegq}
\eeqn

\begin{figure}[t]
    \centering
    \includegraphics[height=10cm]{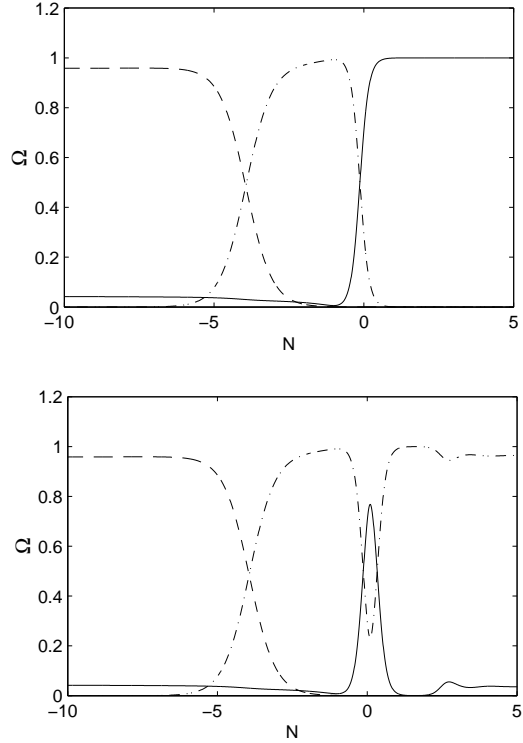}
\caption{Evolution of $\Omega_Q$ (full curve),
$\Omega_{r}$ (dashed curve) and $\Omega_m$ (dotted-dashed curve) for Model I (upper panel), corresponding  to
  permanent vacuum domination and Model II (lower panel), corresponding  to
 temporary vacuum domination. }
\label{fig:omegas}
\end{figure}

Integrating the above field equations, one finds that there are essentially two realistic types of behaviour, illustrated in
Figure~\ref{fig:omegas} (notice that N=-30 corresponds to the Planck epoch; nucleosynthesis occurs around
N=-10, radiation to matter transition around N=-4 and N=0 today).
Model I $(\lambda=9.5, a=0.02, \phi_0=29, \psi_0=15$,  $b=0.002, c = 6 \times 10^{-4}, d=4.5)$  corresponds to the  case where
vacuum domination,  which occurs when  $\Omega_{Q0}>1/2$, is permanent
and Model II $(\lambda=9.5, a=0.1, \phi_0 = 29, \psi_0 = 20$, $b =
0.001, c =8\times 10^{-5}, d = 2.8)$ to  the case where vacuum
domination is transient. Permanent  and transient vacuum domination
have  also been found in the  model of Refs.~\cite{Skordis} and \cite{Barrow}, respectively. Of course, there remains the
(non-realistic) case where accelerated expansion never occurs.

The effective equation  of state for  quintessence  is given
by $p_Q = \omega_{Q} \rho_Q$, where $w_Q = \gamma_Q - 1$, and

\beqa
\gamma_Q \equiv {\rho_Q+p_Q \over  \rho_Q} &= &{\dot\phi^2 + \dot\psi^2
 \over V +
\dot\phi^2/2 + \dot\psi^2/2 }\nonumber\\
 &= &{2(x^2+z^2) \over x^2 + y^2 + z^2}
\label{gama}
\eeqan

A necessary and sufficient condition for the universe to accelerate is
that the deceleration parameter, q, given by

\beq
q = - {a \ddot a\over {\dot a}^2}={1\over 2}  (1+3 w_Q \Omega_Q +
\Omega_r)~,
\label{q}
\eeqn
where $\Omega_r$ is the fractional radiation energy density, is negative.

\begin{figure}[t]
\centering
\includegraphics[height=10cm]{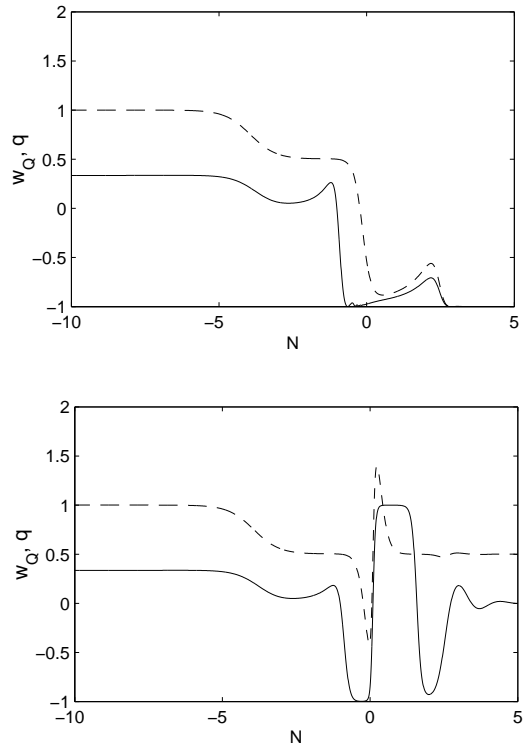}
\caption{  Evolution of the equation
of state parameter $w_Q$ (full curve) and the deceleration parameter $q$
(dashed curve) for Model I (upper
 panel) and Model II (lower panel).}
\label{fig:wandq}
\end{figure}

In Fig.~\ref{fig:wandq}, we show the evolution of $w_Q$ and $q$ Model I
 (uppper panel) and  Model II (lower panel).

In both models, the equation of state has reached $w_Q\simeq -1$ for
the present time, as favored by the available data \cite{efs} (and making
it hard to distinguish from a cosmological constant) but,
whereas in Model I $\omega_Q$  remains negative, in Model II it is
in the process of increasing today towards positive values, then oscillates
 slightly  until it
reaches its 
asymptotic value. Similarly, in both models, the deceleration parameter
is negative today  but, whereas for Model I 
$q$  remains negative, in Model II it oscillates and becomes positive before
it reaches its  asymptotic value.

\begin{figure}[t]
\centering
\includegraphics[height=10cm]{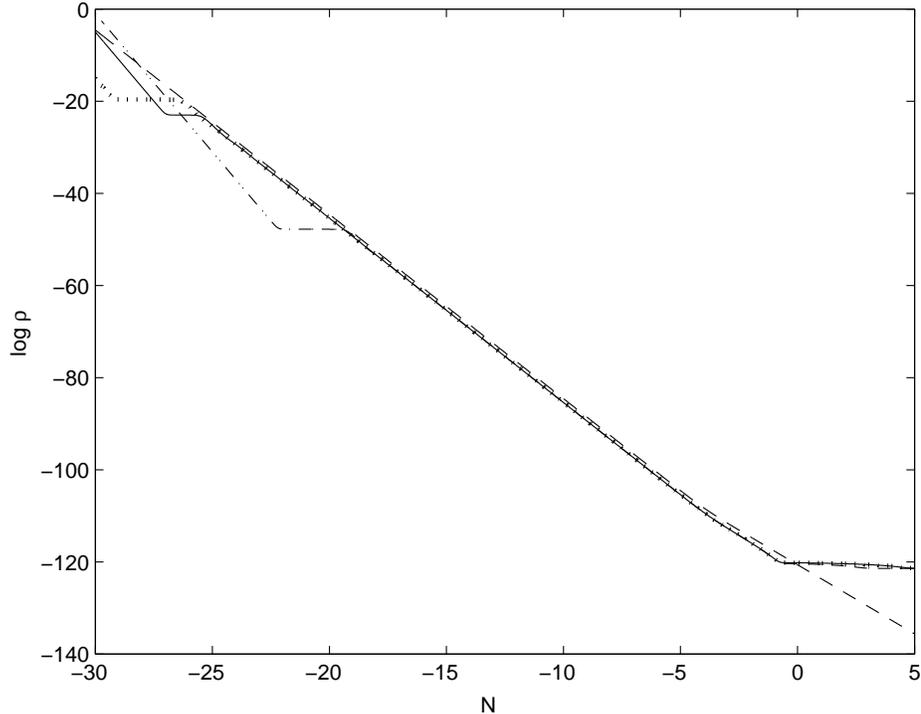}
\caption{Evolution of the background (matter+radiation) energy density
(dashed) and  quintessence energy density, $\rho_Q$, for Model I, for
different initial conditions:   $\rho_{Q}\ll\rho_{r}$    (dotted),
$\rho_{Q}\sim    \rho_{r}$     (solid)    and    $\rho_{Q}>\rho_{r}$
(dotted-dashed).}
\label{fig:rhof}
\end{figure}

Scaling behaviour is apparent in Figure~\ref{fig:rhof},
where we plot quintessence energy density $\rho_Q$ as a function of
the scale factor for Model I, for different initial conditions,
namely,  $\rho_Q\ll \rho_r$,  $\rho_Q>\rho_r$ and  $\rho_Q\sim  \rho_r$.
After showing some initial transient, each solution scales with the
dominant  matter component before $\rho_Q$ begins to dominate. Model II
presents similar behaviour but, as the  scalar fields never stop rolling,
matter-dominated scaling evolution is soon resumed.
Our models seem to be
more sensitive to  changes in  the initial  conditions than models with just one scalar field (this is to be expected since there is more freedom 
e.g. in the way kinetic energy is shared between the two fields) but no fine tuning of the initial conditions is needed.  

\begin{figure}[t]
    \centering
    \includegraphics[height=10cm]{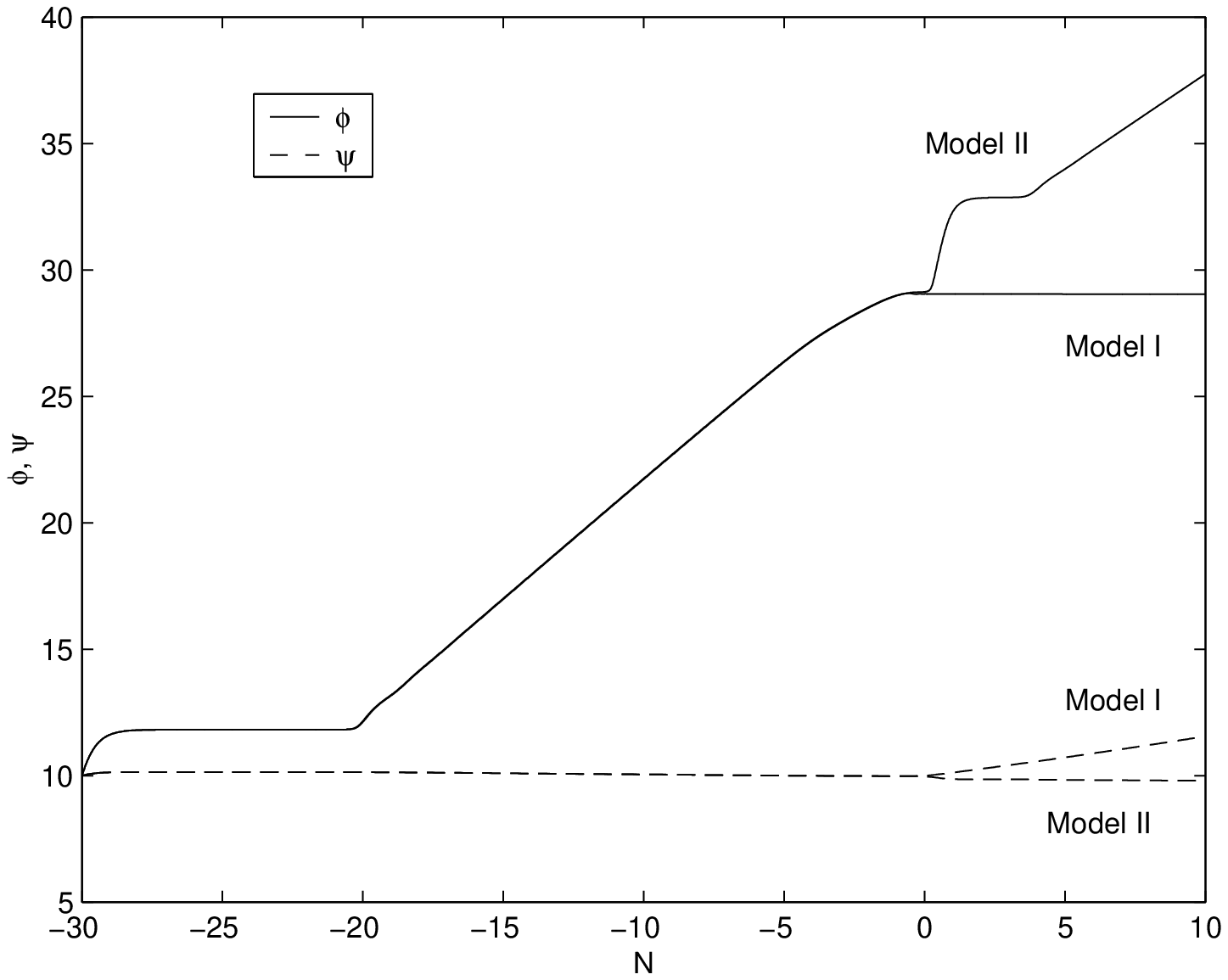}
\caption{Evolution  of  the quintessence  fields,  $\phi$   and
$\psi$, for Models I and II. }
\label{fig:fields}
\end{figure}

Permanent vacuum domination takes place when at least the $\phi$ field
ends up settling  at the minimum of the  potential, thus corresponding
to a cosmological constant;  in  Model I, both fields settle at the minimum of the potential, see Figure~\ref{fig:fields}. Transient vacuum domination occurs either
when the potential has no local minimum or $\phi$ arrives at the local
minimum with enough kinetic energy to roll over the barrier and resume
descending the potential. Notice that the
evolution of $\psi$ is slight compared with $\phi$, 
especially for Model II.

Both models satisfy present bounds on relevant cosmological observables.   
The tightest  bound  comes from  nucleosynthesis, 
$\Omega_Q(N\sim -10) <  0.044$,
requiring $\lambda>9$ \cite{Bean}
from the most recent CMB data, $\Omega_{Q}  < 0.39$ at last scattering, 
is less stringent than the nucleosynthesis bound. Other
bounds we take into account  are: $\Omega_{m} = 0.3 \pm 0.1$, $w_{Q}
<  -  0.6$, $\Omega_{Q}  = 0.65  \pm  0.05$ \cite{Perlmutter1}  and
$h=0.65  \pm   0.05$  \cite{PDG} today.   Indeed,  Model   I  has  $h=0.6$,
$\Omega_{Q} = 0.7$, $\Omega_{m} = 0.3$, $w_{Q}=-1$ and $q=-0.5$
today, $\Omega_{Q}=0.042$ at nucleosynthesis.  Similar  values are found
for Model  II, namely $h=0.6$,  $\Omega_{Q}=0.7$, $\Omega_{m}=0.3$,
$w_{Q}=-0.9$    and   $q=-0.4$    today,    $\Omega_{Q}=0.042$ at
 nucleosynthesis.

We have studied the nature of our solutions for a rather broad range
of parameters of the potential. 
We have found that it is possible to
obtain permanent or transient vacuum domination, satisfying present
bounds on observable cosmological parameters, for various combinations 
of the potential parameters.


\section{ Conclusions}

We have discussed a  potential involving two coupled scalar fields that leads 
 to an interesting
asymptotic quintessence  behaviour and accounts for  data arising from
high redshift  Type Ia Supernovae,  CMB and cluster  mass distribution
data.   The   late  time   dynamics   arising   from  this  potential,
Eq~(\ref{pot}),  is fairly  rich as  it gives  origin to  models where
either the  acceleration regime is  eternal (Model  I)  or  instead a  regime  where acceleration  is
transient (Model II) although a
(non-realistic) late time  non-accelerating dynamics is also possible.
A  relevant  issue of  this  proposal is  that  it  allows evading  the
conclusions of  Refs. \cite{Hellerman,Fischler}, in  what concerns the
stability of a supersymmetric  potential.  
 Furthermore, since, in
Model II,  acceleration is transient  and occurs only at  present, this
model is consistent with the  underlying framework of string theory as
is  does not present  cosmological horizons  that are  associated with
eternally accelerating universes.

We conclude  that the late  time dynamics arising from this two-field
quintessence model is consistent with  the observations as well as the
theoretical  requirements of  stability of  the  supersymmetric ground
state and  the asymptotic behaviour of string  theory states, provided
the observed  accelerated expansion of  the universe is  transient and
decelerated expansion is soon resumed. This solution has been recently
proposed  to solve the  contradiction between  
accelerated  expansion  and  string  theory \cite{Kolda},  on  general
grounds; in this work,  we have presented a concrete example of a 
  two-field model
that   exhibits this desirable feature.



\begin{thebibliography}{99}


\bibitem{cmbexp} P. de Bernardis et al., Nature {\bf 404} (2000) 955;

A.E.  Lange et al., Phys. Rev. {\bf D64} (2001) 042001;

T. Padmanabhan and S.K. Sethi, astro-ph/0010309.

\bibitem{otherexp} N. Bahcall, J.P. Ostriker, S.J.  Perlmutter and P.J.
Steinhardt,  Science {\bf 284} (1999) 1481;

L. Wang, R.R. Caldwell, J.P. Ostriker and P.J. Steinhardt, Astrophys. J. {\bf 
530} (2000) 17, and references therein. 

\bibitem{supern} S.J. Perlmutter  et al.  (The Supernova Cosmology
Project), Ap. J. {\bf 483} (1997) 565; Nature {\bf 391} (1998) 51;

A.G.  Riess  et al., Astron. J. {\bf 116} (1998) 1009.

P.M. Garnavich  et al., Astrophys. J. {\bf 509} (1998) 74.

\bibitem{Ratra}  B. Ratra and  P.J.E.  Peebles,  Phys.\ Rev.\ {\bf  D37} (1988)
3406.
 
 
\bibitem{Caldwell} R.R.  Caldwell, R.  Dave and P.J.  Steinhardt, Phys.\ Rev.\
 Lett.\ {\bf 80} (1998) 1582.
 
\bibitem{Zlatev} I. Zlatev, L. Wang and P.J. Steinhardt, Phys.\ Rev.\ Lett. 
 {\bf 82}
(1999) 986.
 \bibitem{Burgess} C.P. Burgess, R.C. Myers and F. Quevedo, Phys.\ Lett.\ {\bf B495} 
(2000) 384.

 
\bibitem{Linde} A.A. Linde, Phys.\ Lett.\ {\bf B249} (1990) 18.

 
\bibitem{Bento2} M.C. Bento, O. Bertolami and P.M. S\'a, Phys.\ Lett.\ {\bf B262} 
(1991) 11; Mod. Phys. Lett. {\bf A7} (1992) 911.
 

\bibitem{Kofman} L. Kofman, A.A. Linde and A.A. Starobinsky, Phys.\ Rev.\ Lett.  {\bf 76} 
(1996) 1011 ; Phys.\ Rev.\ {\bf D56} (1997) 3258.

 

\bibitem{Bertolami2}  O. Bertolami and G.G. Ross, Phys.\ Lett.\ {\bf B171} 
(1986) 163.


\bibitem{Hellerman} S. Hellerman, N. Kaloper and L. Susskind,
 JHEP {\bf 0106} (2001) 003.


\bibitem{Fischler} W. Fischler, A. Kashani-Poor, R. McNess and S. Paban,
 JHEP {\bf 0107} (2001) 003.

\bibitem{Perlmutter1} S.J. Perlmutter, M.S. Turner and M. White, Phys.\ Rev.\ 
Lett. {\bf 83} 
(1999) 670.

\bibitem{BentoBS} M.C. Bento, O. Bertolami and N.C. Santos, astro-ph/0106405.

\bibitem{Skordis} A. Albrecht and C. Skordis, Phys.\ Rev.\ Lett.\ {\bf 84} (2000) 2076.


\bibitem{Barrow} J.  D. Barrow, R.   Bean and J. Magueijo, Mon. Not. R. Ast. Soc. {\bf 316} (2000) L41.

\bibitem{efs} G. Efstathiou, Mon. Not. R. Ast. Soc. {\bf 3102} (1999)842; 

S. Podariu and B. Ratra, Ap. J {\bf 532} (2000) 109.

\bibitem{Bean} R. Bean, S.H. Hansen and A. Melchiorri, Phys. Rev. {\bf D64} (2001) 103508.


\bibitem{PDG} Particle Data Group, Review of Particle Properties,
The European Phys. J. {\bf C15} (2000) 1.



\bibitem{Kolda} C. Kolda and W. Lahneman, hep-ph/0105300. 



\end{thebibliography}
\end{document}